\documentclass[journal=nalefd,manuscript=letter]{achemso}

\usepackage{graphicx, color}
\usepackage{amsfonts,amssymb,amsmath}
\usepackage{bm}
\usepackage{soul}

\author{Matthias R{\"o}\ss ler}
\author{Dingxun Fan}
\author{Felix M{\"u}nning}
\affiliation[University of Cologne]
{University of Cologne, Physics Institute II, Z\"ulpicher Str. 77, 50937 K\"oln, Germany}
\author{Henry F. Legg}
\affiliation[University of Basel]
{Department of Physics, University of Basel, Klingelbergstrasse 82, CH-4056 Basel, Switzerland}
\author{Andrea~Bliesener}
\affiliation[University of Cologne]
{University of Cologne, Physics Institute II, Z\"ulpicher Str. 77, 50937 K\"oln, Germany}
\author{Gertjan Lippertz}
\affiliation[University of Cologne]
{University of Cologne, Physics Institute II, Z\"ulpicher Str. 77, 50937 K\"oln, Germany}
\alsoaffiliation[University Leuven]
{KU Leuven, Quantum Solid State Physics, Celestijnenlaan 200 D, 3001 Leuven, Belgium}
\author{Anjana Uday}
\author{Roozbeh Yazdanpanah}
\author{Junya Feng}
\author{Alexey A. Taskin}
\affiliation[University of Cologne]
{University of Cologne, Physics Institute II, Z\"ulpicher Str. 77, 50937 K\"oln, Germany}
\author{Yoichi Ando}
\affiliation[University of Cologne]
{University of Cologne, Physics Institute II, Z\"ulpicher Str. 77, 50937 K\"oln, Germany}
\email{ando@ph2.uni-koeln.de}

\title{Top-down fabrication of bulk-insulating topological insulator nanowires for quantum devices}

\begin{document}

\begin{abstract}
In a nanowire (NW) of a three-dimensional topological insulator (TI), the quantum-confinement of topological surface states leads to a peculiar subband structure that is useful for generating Majorana bound states. Top-down fabrication of TINWs from a high-quality thin film would be a scalable technology with great design flexibility, but there has been no report on top-down-fabricated TINWs where the chemical potential can be tuned to the charge neutrality point (CNP). Here we present a top-down fabrication process for bulk-insulating TINWs etched from high-quality (Bi$_{1-x}$Sb$_{x}$)$_2$Te$_3$ thin films without degradation. We show that the chemical potential can be gate-tuned to the CNP and the resistance of the NW presents characteristic oscillations as functions of the gate voltage and the parallel magnetic field, manifesting the TI-subband physics. We further demonstrate the superconducting proximity effect in these TINWs, preparing the groundwork for future devices to investigate Majorana bound states. 
\end{abstract}


In the two-dimensional (2D) Dirac surface states of a three-dimensional (3D) topological insulator (TI), the electron spin axis is dictated by the momentum of the electron, a property called spin-momentum locking \cite{Ando2013}. This feature is useful not only for spintronics \cite{Breunig2022} but also for generating non-Abelian Majorana bound states \cite{Sato2017} by using the superconducting (SC) proximity effect \cite{FuKane2008}, which may open the door for fault-tolerant topological quantum computing \cite{Nayak2008}. For the creation of Majorana bound states, reducing the dimensions of a 3D TI into a nanowire (NW) is useful \cite{Cook2011}. The quantum confinement of the 2D surface states into a NW leads to the appearance of a series of one-dimensional (1D) subbands, offering an interesting platform for various quantum devices \cite{Breunig2022}. To utilize the unique characteristics of these subbands for quantum devices, it is often important to tune the chemical potential to the charge neutrality point (CNP) of the Dirac dispersion \cite{Legg2021, Legg_Diode2022, Legg2022}. This means that a TINW should be bulk-insulating to become useful.

In the past, TINWs have been fabricated by various techniques: Vapour-Liquid-Solid (VLS) growth \cite{Peng2010,Hong2014,  Jauregui2016, Munning2021, ChaRev2019}, exfoliation of a bulk single crystal \cite{Cho2015, Cho2016}, sandwiching between different materials \cite{Bai2022}, selective-area-growth (SAG) \cite{Kolzer2020, Rosenbach2020}, and top-down etching of a 2D film \cite{Ziegler2018, Fischer2022}. 
Among them, the SAG and top-down etching techniques offer a large design flexibility, including the possibility to fabricate curved or branched NWs and NW-circuits, and hence they are particularly appealing as the NW-fabrication techniques for quantum devices. However, there has been no report of a successful tuning of the chemical potential to the CNP in NWs fabricated with these methods. In the case of SAG, it is difficult to selectively grow a TI material, for example (Bi$_{1-x}$Sb$_{x}$)$_2$Te$_3$ (BST) \cite{Zhang2011}, in the bulk-insulating composition \cite{Kolzer2020, Rosenbach2020}, while the top-down etching has been applied only to the TI material HgTe \cite{Ziegler2018, Fischer2022}, in which the CNP is buried in the bulk valence band. In this paper, we report successful fabrication of bulk-insulating TINWs with a top-down etching of BST thin films, and we show that the chemical potential in such NWs can successfully be tuned to the CNP by electrostatic gating. We further demonstrate that these NWs exhibit the peculiar TI-subband physics and can be used for superconducting devices to search for Majorana bound states.


The etched TINW devices have been prepared from bulk-insulating (Bi$_{1-x}$Sb$_{x}$)$_2$Te$_3$ (BST) thin films grown by  molecular-beam epitaxy (MBE), because BST films are suitable for surface-transport studies \cite{Kong2011, Zhang2011, Yang2014, Yoshimi2015, Yang2015, Koirala2015, Taskin2017PHE}. Since air-exposure and heating tend to alter the properties of BST \cite{Lang2012, Ngabonziza2015, BSTox2018}, our films are ex-situ capped with Al$_2$O$_3$ in an atomic-layer-deposition (ALD) machine immediately after the growth (the film is briefly exposed to air during the transfer) and the process temperatures during the device fabrication are limited to $\leqslant$ 120 $^{\circ}$C.

\begin{figure}[t!]
\includegraphics[width=0.89\columnwidth]{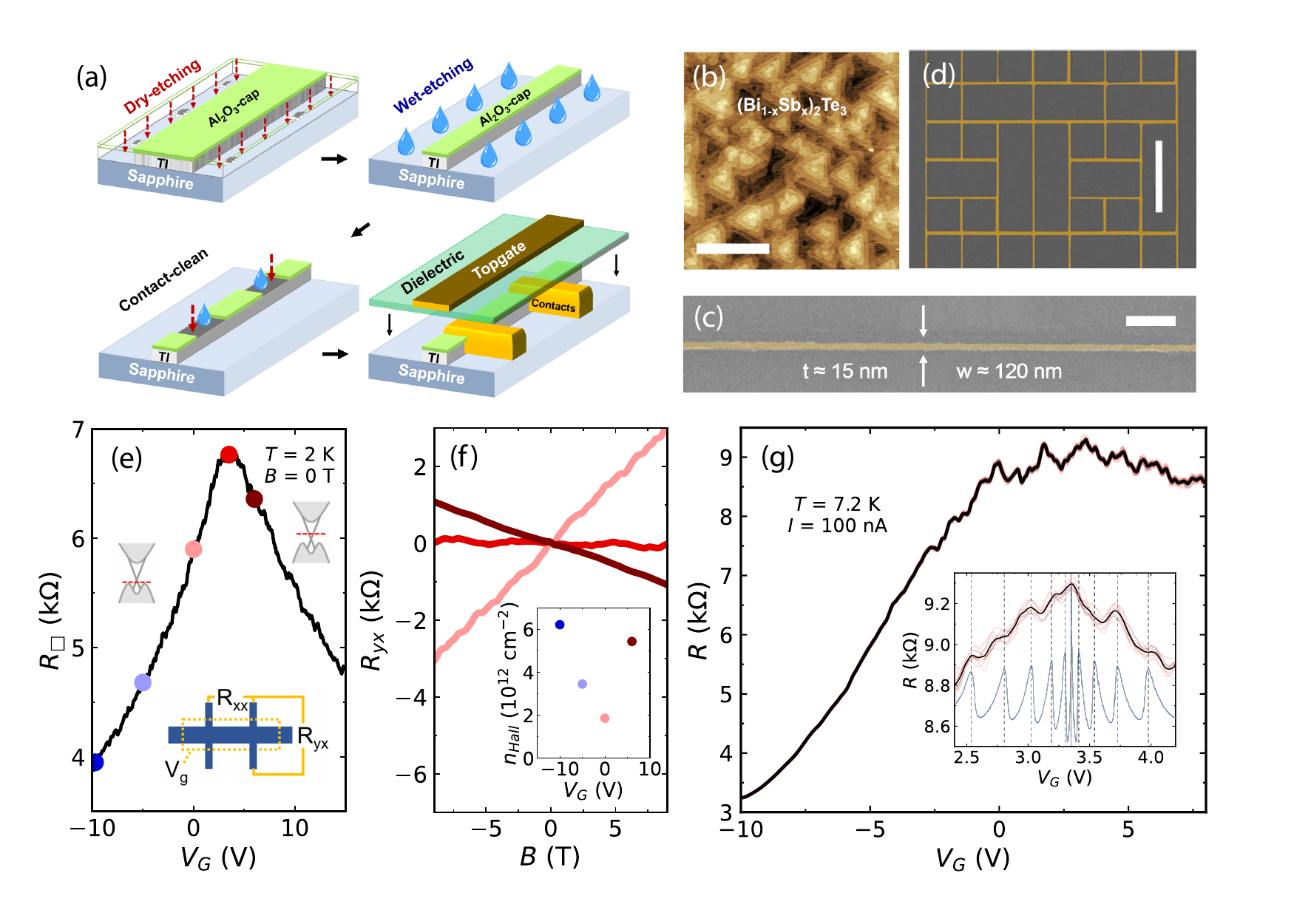}
\caption{\linespread{1.05}\selectfont{}
\textbf{a,} Schematics of the fabrication process: an MBE-grown BST thin film capped with Al$_2$O$_3$ is patterned with EB lithography and physically etched with Ar plasma into nanoribbons; consecutive chemical wet-etching is carried out to obtain NWs with a smooth edge; after ex- and in-situ contact cleaning, Pt/Au or Nb electrodes are sputter-deposited. Some devices are encapsulated by an Al$_2$O$_3$ dielectric and a Pt/Au top gate is fabricated.
\textbf{b,} AFM image of the surface of a BST grown on sapphire with characteristic quintuple-layer terrace steps. Scale bar 1 $\mu$m.
\textbf{c,} SEM image of a long NW with a width of 120 $\pm$ 20 nm and a thickness of 15 $\pm$ 2 nm. Scale bar 1 $\mu$m.
\textbf{d,} SEM image of a NW-network. Scale bar 10 $\mu$m.
\textbf{e,} $V_G$-dependence of the sheet resistance $R_{\square}$ at 2 K of a 500-nm-wide Hall bar prepared with the aforementioned fabrication process.
\textbf{f,} $B$-dependencies of $R_{yx}$ of the same Hall bar at different $V_G$ values that are color-highlighted in panel \textbf{e}; inset shows the carrier density $n_{\rm Hall}$ calculated from the Hall coefficient, with which a mobility of $\sim$600 cm$^{2}$/Vs can be estimated, confirming the absence of degradation in the fabrication process.
\textbf{g,} $V_G$-dependence of the four-terminal resistance $R$ in Device 1 measured between contacts 4--5 (see Supplementary Figure S4) at 7.2 K. Thin red lines show the results of 10 uni-directional $V_G$-sweeps and the thick black line shows their average. Inset shows the $R(V_G)$ behavior near the CNP; dashed vertical lines mark the peak positions expected from theory (blue line) for a gate capacitance $C_G$ = 55 pF obtained from a simple electrostatic model. 
}
\label{fig:Fig1}
\end{figure}

Figure 1a shows a schematic of our top-down fabrication process. After the patterning with electron-beam (EB) lithography, dry-etching with Ar plasma is performed first to remove unnecessary parts of the BST/Al$_2$O$_3$ film. Subsequently, a chemical wet-etching with diluted H$_2$O$_2$/H$_2$SO$_4$ is carried out to remove damaged edges of the remaining BST to obtain NWs with a smooth edge. Next, the contact areas are exposed by employing EB lithography and removing the Al$_2$O$_3$ capping layer in heated Transene Type-D etchant; the exposed BST surface is then cleaned in diluted HCl and further by low-power Ar plasma in the metallization machine. The electrodes (either Pt/Au or Nb) are sputter-deposited in UHV. For our top-gate devices, we deposited a 50-nm-thick Al$_2$O$_3$-dielectric in the ALD machine and then fabricated the top-gate electrode with Pt/Au. Further details are described in the Methods section.

An atomic-force-microscopy (AFM) image of the surface of a typical pristine BST thin film used in our work is shown in Figure 1b. The well-developed terraces with the step height of $\sim$1 nm are a signature of high-quality epitaxial growth. As shown in Figure 1c, etched NWs can have arbitrary channel lengths with only small fluctuations in the width; the NW pictured here had the width $w \simeq$ 120 $\pm$ 20 nm and the thickness $t \simeq$ 15 $\pm$ 2 nm. With such dimensions, the quasi-1D subbands in the NW have an energy spacing of several meV \cite{Cook2012, Munning2021, Breunig2022}. Figure 1d showcases that this approach allows to fabricate advanced geometries and arbitrary complex networks of bulk-insulting TINWs. 

To characterize the properties of the BST film after the NW-fabrication process, we fabricated a relatively wide ``ribbon'' with a width $w$ = 500 nm, for which the Hall measurement is possible (note, in contrast, the Hall voltage would not appear in a strictly 1D system, and hence the interpretation of the Hall measurement is difficult in a quasi-1D system). The sheet-resistance $R_{\square}$ of such a ribbon sample at 2 K is shown in Figure 1e as a function of the gate voltage $V_G$, and Figure 1f shows the magnetic-field dependence of the Hall resistance $R_{yx}$ of this ribbon sample for three $V_G$ values. The peak in $R_{\square}(V_G)$ corroborated with a sign change in $R_{yx}$ gives clear evidence for the CNP-crossing and the bulk-insulating nature of the ribbon after the NW-fabrication process. 
The $R_{\square}$ and $R_{yx}$ data for $V_G$ = 0 V give the hole density $p$ = $1.8 \times 10^{12}$ cm$^{-2}$ and the mobility $\mu$ = 600 cm$^{2}$/Vs, which are close to the values of the pristine material (Supplementary Figure S2). These results demonstrate that our NW-fabrication process causes little degradation in bulk-insulating BST.

Once the width of the ribbon is reduced to the ``nanowire'' regime, which is typically realized for $w \lesssim$ 250 nm, the subband physics starts to show up \cite{Munning2021, Breunig2022}. One characteristic feature is the $V_G$-dependent oscillation of the NW resistance $R$.  Such a behavior observed in a $w \simeq$ 150 nm NW (the device picture is shown in Supplementary Figure S4) is presented in Figure 1g, where thin red lines show the results of 10 uni-directional $V_G$-sweeps and the thick black line shows their average. One can see that the pattern of the oscillations is disordered but is essentially reproducible. As shown in the inset of Figure 1g for a smaller $V_G$ range near the CNP, reproducible maxima and minima with a characteristic quadratic spacing can be identified. This feature has been elucidated in VLS-grown NWs \cite{Munning2021} to stem from the regular change in the density of states that occurs when the chemical potential crosses the edge of the quantum-confined subbands. The expected positions of the maxima in $R(V_G)$ can be calculated by using the gate capacitance $C_G$ = 55 pF/m estimated from a simple electrostatic model (see Supporting Information), and the theoretically expected positions are in reasonable agreement with the experimentally observed peaks near the CNP. For this NW, we have also estimated the mobility $\mu$ from the $R(V_G)$ data and found no evidence for degradation (see Supporting Information).


\begin{figure}[t]
\includegraphics[width=1\columnwidth]{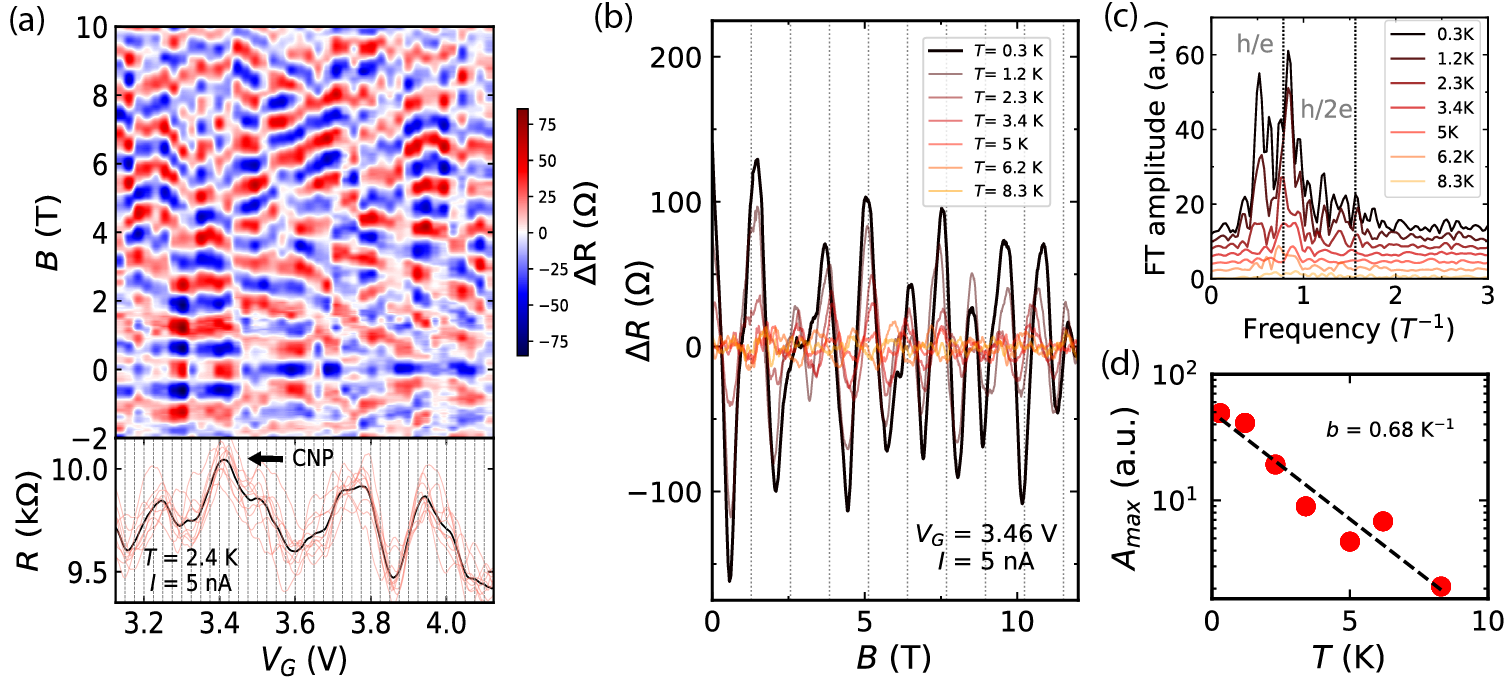}
\caption{\linespread{1.05}\selectfont{}
\textbf{a,} Color mapping of the resistance change $\Delta R$ in Device 1 at 2.4 K as a function of $V_G$ and the parallel magnetic field $B$, presenting occasional $\pi$-phase shifts in the AB oscillations. Lower panel shows the $R(V_G)$ behavior at $B$ = 0 in the same $V_G$ range, with dashed lines indicating the positions of individual magnetic-field sweeps.
\textbf{b,} AB oscillations at various temperatures from 0.3 to 8.3 K measured near the CNP at $V_G$ = 3.46 V.
\textbf{c,} Fourier transform (FT) of the $\Delta R(B)$ data shown in \textbf{b}. Dotted vertical lines indicate the frequencies corresponding to $\Phi_0 = h/e$ and $\Phi_0 /2$.
\textbf{d,} Exponential decay of the FT amplitude $A_{max}$ with $T$ obtained from the data in \textbf{c}.
}\label{fig:Fig2}
\end{figure}

It is well known \cite{Peng2010, Xiu2011, Hong2014, Cho2015, Jauregui2016, Ziegler2018, Kim2020_AB, Rosenbach2020} that the quantum-confined subbands in TINWs also give rise to oscillations of $R$ in parallel magnetic fields with the period of the flux quantum $\Phi_0=h/e$. This phenomenon is called Aharonov-Bohm (AB) oscillations, because it stems from the AB phase affecting the quantized angular momentum of the subbands \cite{Zhang2010, Bardarson2010}. Since the application of $V_G$ leads to oscillations in $R$ as well, the AB oscillations in TINWs are known to present $V_G$-dependent $\pi$-phase shifts \cite{Cho2015, Jauregui2016, Sacksteder2016, Kim2020_AB}. 
To observe this interplay between $V_G$ and $B$, we applied a magnetic field $B$ parallel to the NW at various $V_G$ values changed with small steps, and traced the dependence of $R$ on $B$ and $V_G$. Figure 2a (upper panel) shows a color mapping of $\Delta R$ in the plane of $B$ vs $V_G$, where $\Delta R \equiv R(B)-f(B)$ with $f(B)$ a smooth background; the lower panel presents the $R(V_G)$ behavior at $B$ = 0 T for the same gating range, showing that several subband crossings are occurring in this $V_G$ range. 
Although the periodicity in $B$ is often distorted in Figure 2a, the approximate period of the $B$-dependent oscillations, $\sim$1.3 T, is consistent with $\Phi_0$ for the cross-section of this NW (150 $\times$ 30 nm$^2$). One can see that sudden $\pi$-phase shifts occur at several $V_G$ values, and the occurrence of the sudden phase shifts seen in the upper panel is roughly correlated with the peak-dip features in the lower panel (see Supplementary Figure S6 for additional plots). Ideally, a regular checker-board pattern (Supplementary Figure S5) is expected in the plane of $B$ vs $V_G$; however, experimentally the pattern is always distorted \cite{Cho2015, Jauregui2016, Ziegler2018, Kim2020_AB} which may stem from orbital effects, subband splitting due to gating \cite{Legg2022}, or disorder \cite{Sacksteder2016}.

In our NW, the decay of the AB oscillations with increasing temperature is exponential \cite{Dufouleur2013, Jauregui2016, Kim2020_AB} rather than a power law \cite{Peng2010}, suggesting an approximate phase coherence around the perimeter of the NW \cite{Dufouleur2013, Jauregui2016, Kim2020_AB}. Figure 2b shows $\Delta R(B)$ measured in the temperature range of 0.3--8.3 K near CNP at $V_G$ = 3.46 V, and the Fourier transform (FT) of these data are shown in Figure 2c. One can see that the dominant oscillation frequency corresponds to the AB oscillations with the period $\Phi_0 = h/e$, and the Altshuler-Aronov-Spivak (AAS) oscillations \cite{Altshuler1981} with period $h/2e$ is essentially absent \cite{Peng2010}, which is typical for a short TINW \cite{Kim2020_AB}. Figure 2d demonstrates that the decay of the maximum FT amplitude $A_{\rm max}$ with temperature is well described by an exponential behavior, $A_{\rm max} \sim \exp (-bT)$, which has been interpreted to originate from the disorder-induced thermal length $L_T$ that changes as $\sim T^{-1}$ \cite{Dufouleur2013}. By putting $\exp (-bT) = \exp (L_{\rm p}/L_T)$ with $L_{\rm p}$ the perimeter length (= 360 nm in the present NW), $b$ = 0.68 K$^{-1}$ obtained in our analysis corresponds to $L_T \approx$ 0.53 $\mu$m at 1 K, which is shorter than that in VLS-grown TINWs \cite{Dufouleur2013, Jauregui2016, Kim2020_AB}.


\begin{figure}[t]
\includegraphics[width=\columnwidth]{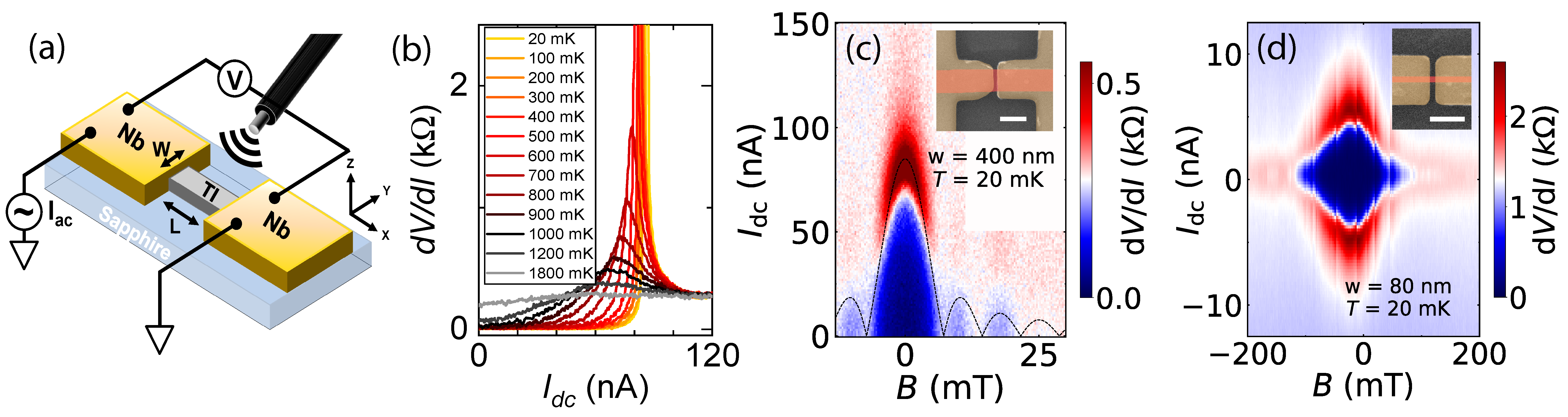}
\caption{\linespread{1.05}\selectfont{}
\textbf{a,} Schematics of the measurements of our TI-based JJ. 
\textbf{b,} Differential resistance d$V$/d$I$ as a function of dc current $I_{\rm dc}$ at various temperatures for Junction 1 made on a 400-nm-wide TI ribbon.
\textbf{c,} Color mapping of d$V$/d$I$ in the $I_{\rm dc}$-vs-$B$ plane for Junction 1 at 20 mK, presenting a Fraunhofer pattern; dashed black line shows the theoretically expected pattern for a conventional junction. Inset shows a false-color SEM-image of Junction 1 with TI (red) and Nb (yellow) on sapphire.
\textbf{d,} Similar mapping of d$V$/d$I$ in the $I_{\rm dc}$-vs-$B$ plane for Junction 2 made on a 80-nm-wide TINW, showing the absence of a Fraunhofer pattern. Inset shows a false-color SEM-image of Junction 2. Scale bars indicate 500 nm.
}\label{fig:Fig3}
\end{figure}

To see if our top-down fabricated TINWs can be a viable platform to search for Majorana bound states \cite{Cook2011, Legg2021}, we have fabricated Josephson junctions (JJs) consisting of Nb electrodes contacting the NW with a small separation ($< 100$ nm). To ensure bulk-insulation in the NW, BST films with a particularly low carrier concentration were chosen for the NW-fabrication (Supplementary Figures S7 and S8). The JJ devices, schematically depicted in Figure 3a, are measured in a quasi-four-terminal configuration and can be exposed to rf radiation to study their Shapiro response.

Figure 3b shows that in our Junction 1, which had a $w$ = 400 nm TI ribbon (see the SEM-image in Figure 3c inset), a well-developed critical current $I_c$ is observed at 20 mK but $I_c$ diminishes at $\sim$1 K, and at 1.8 K there is almost no sign of superconductivity. The temperature dependence of $I_c$ allows for an estimate of the interface transparency of the JJ, as was done in similar JJs \cite{Schueffelgen2019, Bai2022}, which gives a relatively low transparency $\mathcal{T}$ of $\sim$0.5 (see Supporting Information). A different method of estimating the interface transparency based on the excess current observed in the $I$-$V$ curve \cite{Flensberg1988} also gives $\mathcal{T} \approx 0.5$  (see Supporting Information). Hence, there is room for improving the JJ properties by preparing a more transparent interface, which we leave for future works.

When a perpendicular magnetic field $B$ is applied to a usual JJ, the creation of Josephson vortices in the junction leads to the appearance of a Fraunhofer pattern in the $B$-dependence of $I_c$ \cite{GrossReview2016}. Hence, the Fraunhofer pattern is an indication of the phase winding along the width of the junction, which is not expected when the JJ is formed on a NW having no degree of freedom along the width of the junction. Indeed, whereas the Junction 1 made on a 400-nm-wide ribbon shows a Fraunhofer pattern (Figure 3c), the Junction 2 made on a 80-nm-wide NW just presents a monotonic decay of $I_c$ with $B$ (Figure 3d). To the best of our knowledge, our Junction 2 is the first successful JJ made on a TINW having such a narrow width of 80 nm. For Junction 1, the effective area $A_{\rm eff} = 2.3 \times 10^{13}$ m$^2$ indicated by the periodicity of the Fraunhofer pattern (9 mT) is consistent with the geometrical area considering the flux focusing effect \cite{Ghatak2018,Rosenbach2021}.


\begin{figure}[t]
\includegraphics[width=0.95\columnwidth]{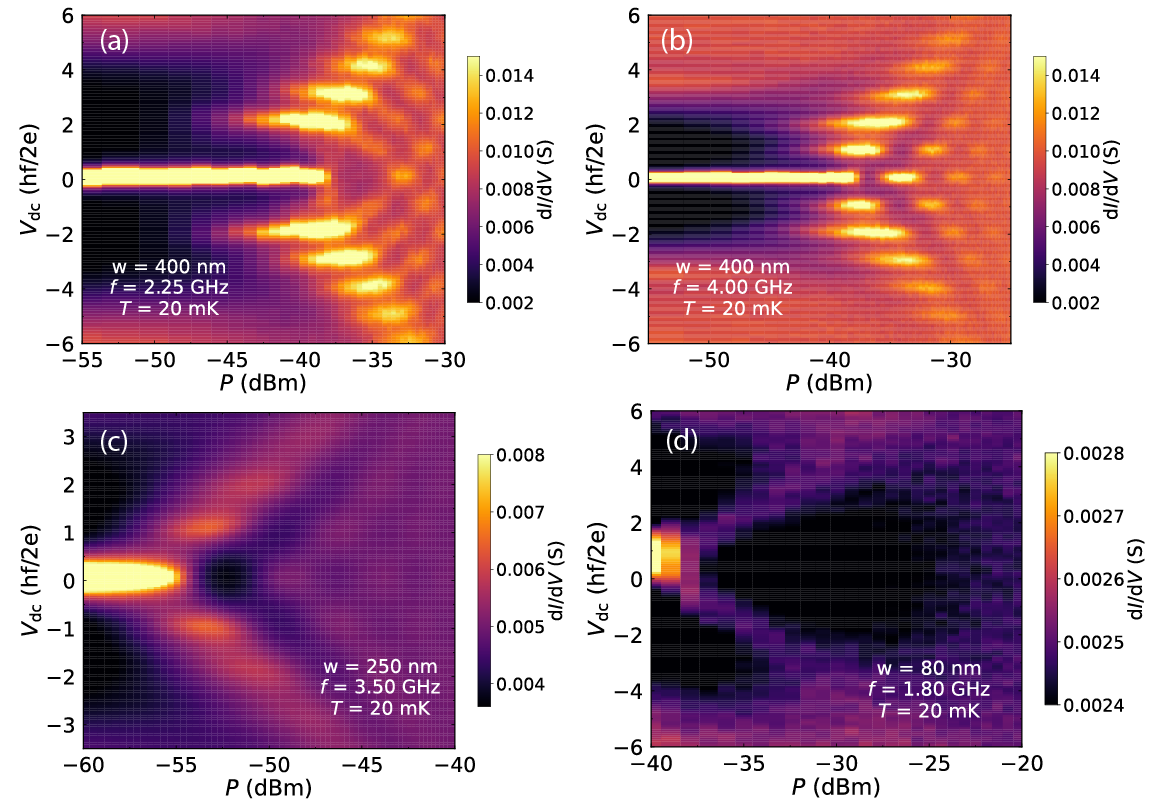}
\caption{\linespread{1.05}\selectfont{}
Color mappings of the differential conductance d$I$/d$V$ at 20 mK in 0 T as a function of rf power $P$ and dc bias $V_{\rm dc}$ to visualize the Shapiro steps. The data on Junction 1 ($w$ = 400 nm) for $f$ = 2.25 GHz and 4 GHz are shown in \textbf{a} and \textbf{b}, respectively; the $n$ = 1 Shapiro step is missing in \textbf{a}, while it is present in \textbf{b}.
The data on Junction 3 ($w$ = 250 nm) at $f$ = 3.5 GHz is shown in \textbf{c}, where the $n$ = 1 Shapiro step is visible. 
The data on Junction 2 ($w$ = 80 nm) at $f$ = 1.8 GHz is shown in \textbf{d}, where no Shapiro step is visible. 
}\label{fig:Fig4}
\end{figure}

We have also studied the behavior of the Shapiro steps in our JJs under rf irradiation. 
Figure \ref{fig:Fig4} shows color mappings of the differential conductance d$I$/d$V$ measured in three JJs as a function of the rf power $P$ and the applied dc bias $V_{\rm dc}$ normalized by $\frac{hf}{2e}$ with $f$ the rf frequency; in this type of plots, the horizontal lines appearing at $V_{\rm dc} = n\frac{hf}{2e}$ correspond to Shapiro steps with index $n$.
Figures \ref{fig:Fig4}a and \ref{fig:Fig4}b show the Shapiro response of Junction 1 (made on a 400-nm-wide TI ribbon) at 20 mK for 2.25 and 4 GHz, respectively. The 2.25-GHz data lack the $n$ = 1 step, while the 4-GHz data present all integer steps. At an intermediate frequency of 3 GHz, the 1st Shapiro step was partially observed (Supplementary Figure S11). The missing 1st Shapiro step has been reported before for TI-based JJs \cite{Wiedenmann2016, Bocquillon2017, Schueffelgen2019, Calvez2019, Ronde2020, Rosenbach2021, Fischer2022} and was interpreted to be possible evidence for Majorana bound states which endow $4\pi$-periodicity to the JJ \cite{Fu2008}. However, the Shapiro response alone can never give conclusive evidence for Majorana bound states because other mechanisms of nontopological origin can also cause the 1st Shapiro step to become missing \cite{Chiu2019, Dartiailh2021, Zhang2022, Fischer2022}. Hence, our observation just demonstrates that our top-down fabrication process does not adversely affect the SC proximity effect in TIs and can be useful for future studies of Majorana bound states.
 
If one interprets the missing 1st Shapiro step to be due to a $4\pi$-periodic component in the Josephson response \cite{Dominguez2017}, the characteristic crossover frequency $f_{4 \pi}$ for the appearance of the 1st Shapiro step can be linked to the amplitude of the $4 \pi$-periodic current $I_{4 \pi}$ via $f_{4 \pi} = 2 e R_N I_{4 \pi} / h$ with $R_N$ the normal-state resistance of the junction \cite{Wiedenmann2016}. For our Junction 1, we infer $f_{4 \pi} \approx$ 3 GHz, which gives $I_{4 \pi} \approx$ 26 nA and $I_{4 \pi} / I_{\mathrm{c}} \approx$ 0.36.


Of particular interest is how such a Shapiro response changes when the TI-part is narrowed into the NW regime \cite{Laroche2019, Ueda2020, Rosenbach2021, Fischer2022}, but our Shapiro-measurement setup was not good enough for this purpose. For our Junction 3 ($w$ = 250 nm), 3.5 GHz was the lowest $f$ at which we could observe Shapiro steps, and the $n$ = 1 step was present (Figure \ref{fig:Fig4}c). In Junction 2 which had the narrowest NW ($w$ = 80 nm), we detected no Shapiro response at any frequency (Figure \ref{fig:Fig4}d). This result tells us that the electromagnetic coupling between the TINW and the microwave antenna should be improved, which is beyond the scope of this study. Once a stronger coupling is achieved, it would be interesting to clarify the effect of parallel magnetic fields on the TINW-based JJ to detect the possible generation of Majorana bound states \cite{Cook2011, Fischer2022}. Trying to detect the microwave emission from a voltage-biased JJ \cite{Deacon2017,Kamata2018} would also be a promising future direction for proximitized TINWs.


In conclusion, we established a top-down fabrication process for bulk-insulating TINWs that are etched from MBE-grown (Bi$_{1-x}$Sb$_{x}$)$_2$Te$_3$ thin films while maintaining the pristine material's properties. In normal-state transport experiments, characteristic resistance oscillations as a function of both gate voltage and parallel magnetic field are observed, giving evidence that the characteristic TI-subband physics manifests itself in our NWs. We further demonstrate that it is possible to study the superconducting proximity effect in these TINWs by fabricating Josephson junctions with ex-situ prepared Nb contacts, providing the foundation for future devices to investigate Majorana bound states. The process established here is based on widely available clean-room technologies, benefits from the well-controlled properties of high-quality MBE-grown thin films, and enables flexible design of NW geometries. Hence, the results presented here will widen the opportunities for research using mesoscopic topological devices.


\section{Methods}

\subsection{Material growth and device fabrication}
The (Bi$_{1-x}$Sb$_{x}$)$_2$Te$_3$ thin films were grown on sapphire (0001) substrates by co-evaporation of high-purity Bi, Sb, and Te in an ultra-high-vacuum MBE chamber. The flux of Bi and Sb was optimized to obtain the most bulk-insulating films which was achieved with a ratio of 1:6.8. The average thicknesses of the thin films, measured with AFM, were nominally in the range of 15--30 nm. Within minutes after taking the films out of the MBE chamber, they were capped with a 3-nm-thick Al$_2$O$_3$ layer grown by atomic layer deposition (ALD) at 80 ${\rm ^o C}$ using a Ultratec Savannah S200. The carrier density and the mobility of the pristine films were extracted from Hall measurements performed at 2 K using a Quantum Design PPMS.
Gate-tunable multi-terminal NW devices were fabricated using the top-down process, the outline of which was described in the main text. After defining the nanowire shape with EB lithography, the film was first dry-etched using low-power Ar plasma. This process can degrade the TI-side-surfaces and can leave residual hard-to-remove resist patches behind. Therefore, a (H$_2$SO$_4$:H$_2$O$_2$):H$_2$O (1:3):8 aqueous solution was used to remove the residual resist as well as the damaged material from the sidewalls of the NWs. To ensure a more controlled etch, the etchant and DI water were cooled to $\sim$8 $^{\circ}$C. A second step of EB lithography was employed to form two kinds of contacts. Normal-state measurements were made with metal contacts that connect to the remaining BST-leads next to the NW to have an enlarged contact area and to avoid unintentional contact-doping in the NW. On the other hand, superconducting contacts were placed in a regular Josephson junction layout directly on top of the NW. In the contact areas, the Al$_2$O$_3$ capping layer was removed by aluminum etchant (Transene Type-D) heated to 50 $^{\circ}$ C. After ex-situ and in-situ contact cleaning in diluted HCl:H$_2$O (1:10) and low-power Ar-plasma in the sputtering machine just before metallization, 5/45 nm Pt/Au contacts or 45/5 nm Nb/Au contacts were deposited by UHV-sputtering. Non-superconducting devices were capped with a 50-nm-thick Al$_2$O$_3$ dielectric grown by ALD at 80 $^{\circ}$C, after which a 5/40 nm Pt/Au top gate was sputter-deposited. In devices with superconducting contacts, the quality of the induced gap was very sensitive to the interface conditions, and for that reason, all the steps between the removal of the Al$_2$O$_3$ capping layer and the contact deposition were carried out in a N$_2$-atmosphere using a Sigma-Aldrich Atmosbag to minimize oxide formation from exposure to air \cite{BSTox2018}. The final dimensions of the devices were determined with scanning electron microscopy.

\subsection{Transport measurements}
Non-superconducting transport measurements were performed in a $^{3}$He cryostat (Oxford Instrument Heliox VL), a system that can be operated in a temperature range from 0.25 K to $\sim$30 K and in magnetic fields up to 13 T. A top-mounted RC-filter box is used to reduce the electromagnetic noise. Superconducting transport measurements are carried out in a dry dilution refrigerator (Oxford Instruments TRITON 300) equipped with a 3-axis vector magnet. In this system, low temperature electronic filtering includes microwave filters, low-pass resistor-capacitor filters and copper-powder filters to reduce electron heating by high-frequency noise. Voltages were measured in a standard four-terminal configuration with a low-frequency lock-in technique with $\sim$13.37 Hz using lock-in amplifiers (NF Corporation LI5645, Stanford Research Systems SR530 and SR830). To apply dc gate voltages or dc bias currents, a Keithley 2450 in combination with a 1-G$\Omega$ resistor was used. For $dV/dI$ measurements, a small ac bias current was generated by using the output voltage of a lock-in amplifier in combination with a 10-M$\Omega$ resistor and added on top of a larger dc current bias. The resulting differential voltage is measured with a lock-in technique.

\section{Author Information}

\subsection{ORCID}
Matthias Rößler: 0000-0001-9911-2835 \\
Felix Münning: 0000-0003-0193-3542 \\
Alexey A. Taskin: 0000-0002-2731-761X \\
Gertjan Lippertz: 0000-0002-4061-7027 \\
Anjana Uday: 0000-0002-9719-0458 \\
Henry F. Legg: 0000-0003-0400-5370 \\
Yoichi Ando: 0000-0002-3553-3355


\subsection{Notes}
The authors declare no competing financial interest.

\begin{acknowledgement}
We are grateful for valuable discussions and support at various stages of this work from A. Rosch, E. Bocquillon, D. Rosenbach, O. Breunig, J. Feng, J. Schluck, N. Zapata, L. Hamdan and T. Zent. This project has received funding from the European Research Council (ERC) under the European Union's Horizon 2020 research and innovation programme (grant agreement No 741121 [Y.A.]. It was also funded by the Deutsche Forschungsgemeinschaft (DFG, German Research Foundation) under CRC 1238 - 277146847 (Subprojects A04 and B01) [Y.A., A.T.] as well as under Germany's Excellence Strategy - Cluster of Excellence Matter and Light for Quantum Computing (ML4Q) EXC 2004/1 - 390534769 [Y.A.]. This work was also supported by the Georg H. Endress Foundation [H.F.L.]. G.L. acknowledges the support by the KU Leuven BOF and Research Foundation Flanders (FWO, Belgium), file No. 27531 and No. 52751.
\end{acknowledgement}

\begin{suppinfo}
The Supporting Information is available free of charge at ....

Properties of BST before and after the NW-fabrication process, Gate-tunable nanowire device image and wiring, Gate-voltage dependent resistance oscillations in TINW, Estimation of the mobility $\mu$ in nanowires, $\pi$-phase shift in the Aharonov-Bohm oscillations, Properties of the pristine BST films used for the Josephson junction devices, Temperature dependence of the critical current in Junction 1, Characteristic parameters of the Josephson-junction devices, Shapiro step data for Junction 1 at $f$ = 3 GHz.

\end{suppinfo}

\begin{flushleft}
{\bf TOC image:}
\end{flushleft}
\vspace{-5mm}
\begin{figure}
\includegraphics[width=0.75\linewidth]{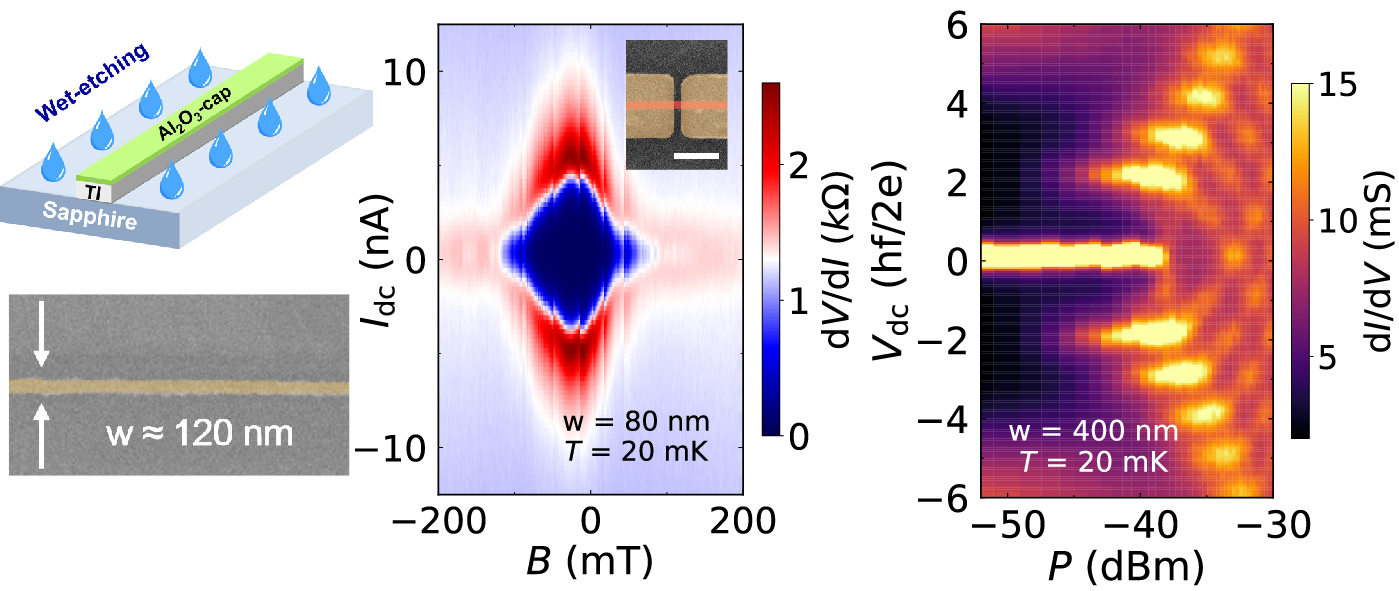}
\end{figure}


\providecommand{\latin}[1]{#1}
\makeatletter
\providecommand{\doi}
  {\begingroup\let\do\@makeother\dospecials
  \catcode`\{=1 \catcode`\}=2 \doi@aux}
\providecommand{\doi@aux}[1]{\endgroup\texttt{#1}}
\makeatother
\providecommand*\mcitethebibliography{\thebibliography}
\csname @ifundefined\endcsname{endmcitethebibliography}
  {\let\endmcitethebibliography\endthebibliography}{}


\end{document}


\setstretch{1}
\setcounter{page}{1}

\section{Properties of BST before and after the NW-fabrication process}

\begin{figure}[h]
\includegraphics[width=0.8\columnwidth]{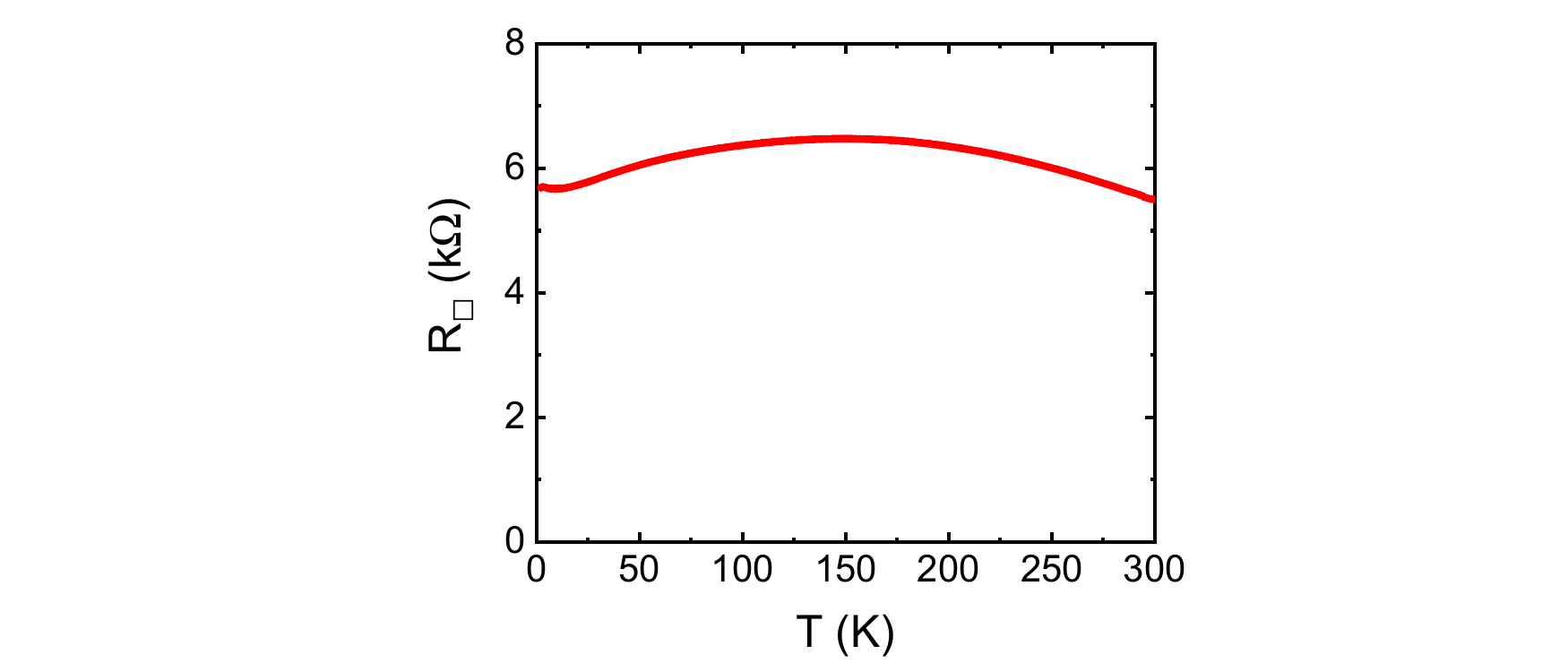}
\caption{\linespread{1.05}\selectfont{}
\textbf{a,} Temperature dependence of the sheet resistance $R_{\square}$ of the 500-nm-wide Hall-bar device discussed in the main text.
}\label{fig:FigS1}
\end{figure}

Figure \ref{fig:FigS1} shows the four-terminal sheet resistance $R_{\square}$ as a function of temperature $T$ in the 500-nm-wide Hall-bar device discussed in the main text (Figs. 1e and 1f), showing the behavior typical for bulk-insulating (Bi$_{1-x}$Sb$_{x}$)$_2$Te$_3$ (BST) thin films \cite{Zhang2011,Yang2015}. 

\begin{figure}[h]
\includegraphics[width=0.8\columnwidth]{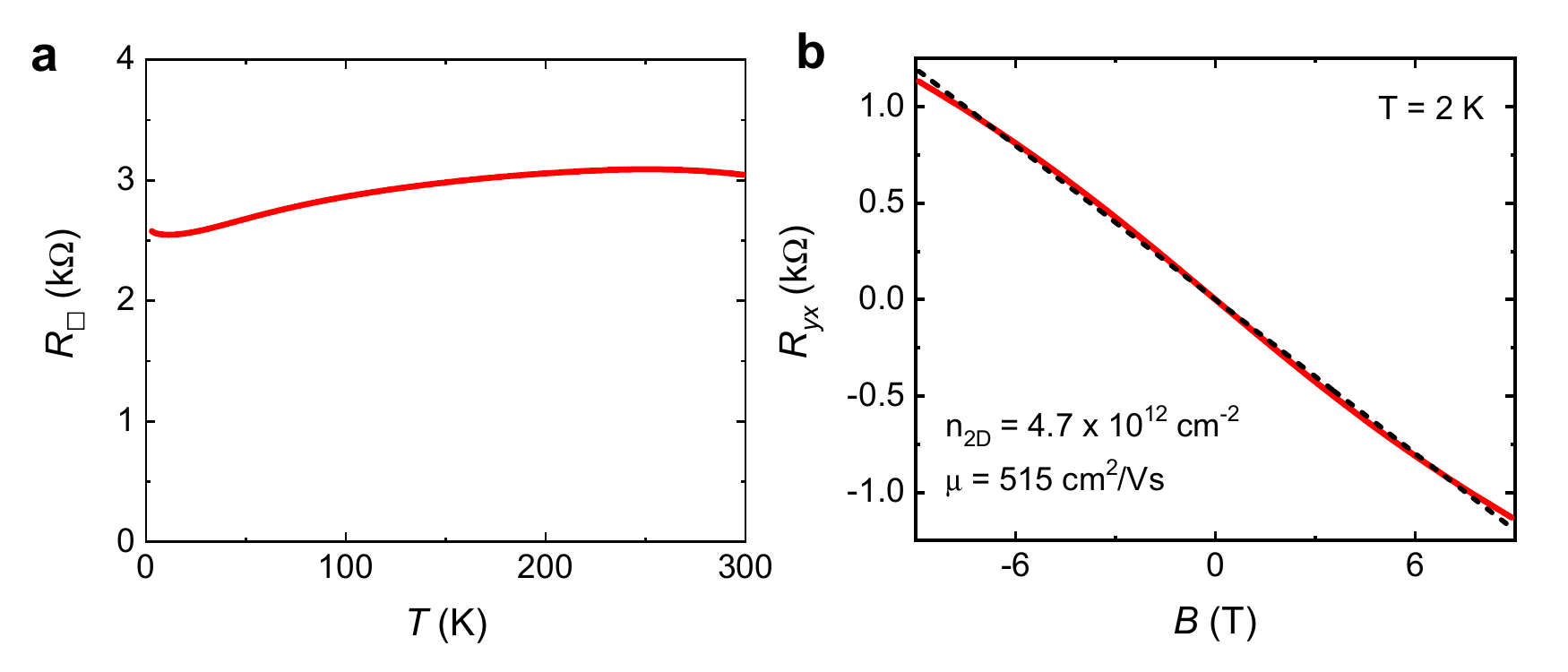}
\caption{\linespread{1.05}\selectfont{}
\textbf{a,} Temperature dependence of the sheet resistance $R_{\square}$ of the pristine BST film used for the 500-nm-wide Hall-bar device discussed in the main text.
\textbf{b,} Hall resistance $R_{yx}$ as a function of magnetic field $B$ of the same thin film. 
}\label{fig:FigS2}
\end{figure}

Figure \ref{fig:FigS2}a shows the $R_{\square}(T)$ behavior of the pristine BST thin film used for this Hall-bar device, and Fig. \ref{fig:FigS2}b shows the Hall resistance $R_{yx}$ of this film as a function of the magnetic field $B$, which gives an electron density $n_{\rm 2D}$ = 4.7$\times10^{12}$ cm$^{-2}$. From these data, we extract the mobility $\mu$ = 515 cm$^{2}$/Vs for the pristine film. This mobility is close to that obtained for the Hall-bar device (600 cm$^{2}$/Vs) after carrying out the fabrication process, indicating that there is essentially no degradation in material's quality.    

\begin{figure}[h]
\includegraphics[width=0.8\columnwidth]{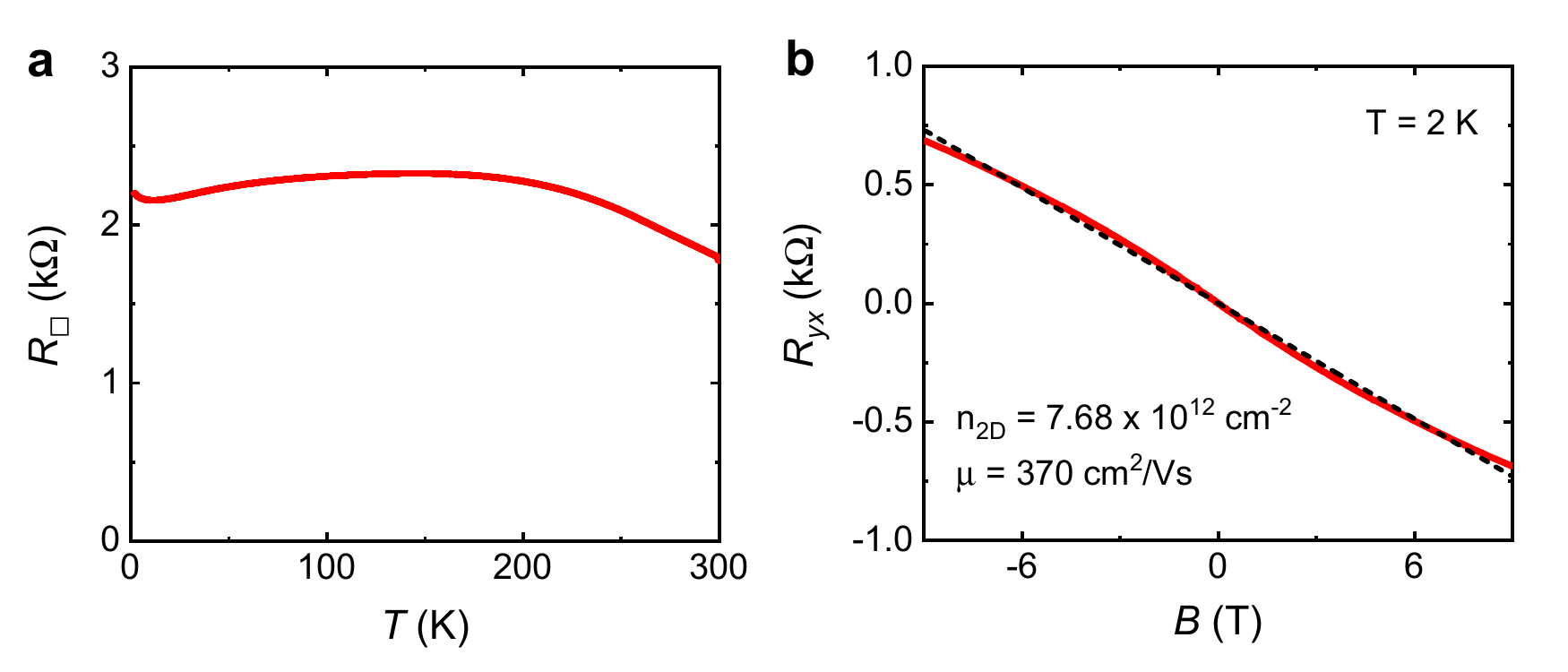}
\caption{\linespread{1.05}\selectfont{}
Temperature dependence of the sheet resistance $R_{\square}$ of the pristine BST film used for the gate-tunable Device 1 used for Fig. 1g and Fig. 2 of the main text.
\textbf{b,} Hall resistance $R_{yx}$ as a function of magnetic field $B$ of the same thin film. 
}\label{fig:FigS3}
\end{figure}

Figures \ref{fig:FigS3}a and \ref{fig:FigS3}b show the $R_{\square}(T)$ behavior and the $R_{yx}(B)$ behavior, respectively, of the pristine BST thin film used for the gate-tunable nanowire (NW) Device 1 before fabrication. These data indicate that the pristine film was a typical bulk-insulating BST film with $n_{\rm 2D}$ = 7.7$\times10^{12}$ cm$^{-2}$ and $\mu$ = 370 cm$^{2}$/Vs. 

Note that the chemical potential tends to shift slightly towards the hole-doped side after the fabrication process, resulting in the 500-nm-wide Hall-bar device and the NW Device 1 to become $p$-type after the fabrication compared to the initially $n$-type pristine films.

\section{Gate-tunable nanowire device image and wiring}

\begin{figure}[h]
\includegraphics[width=0.8\columnwidth]{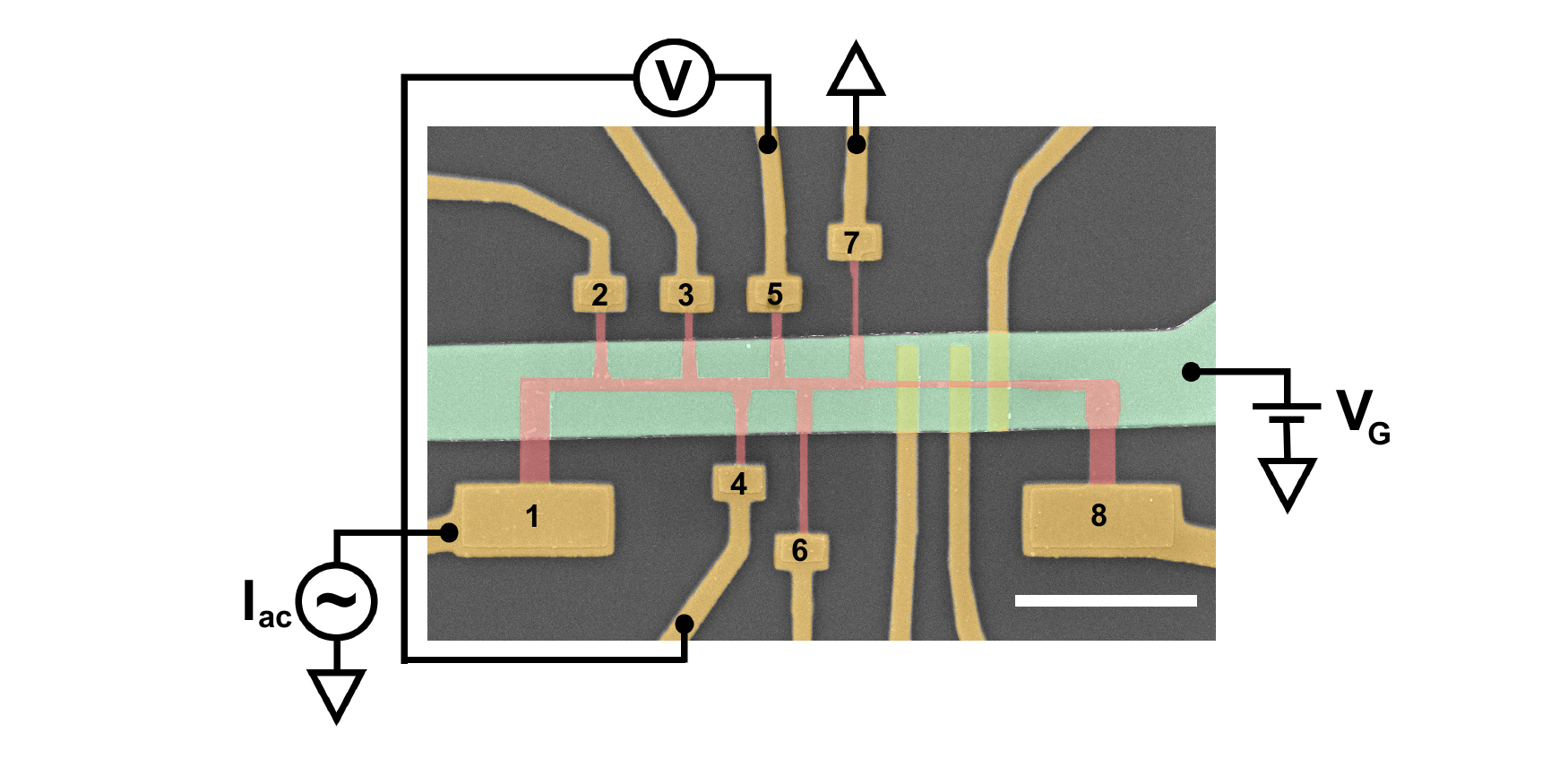}
\caption{\linespread{1.05}\selectfont{}
False-colour SEM image of Device 1 with schematics of the electrical wiring. The Pt/Au leads are colored in dark yellow, the TINW  in red, and the top-gate electrode in green. The four-terminal resistance of the NW was measured between the voltage contact pair 4–5. Scale-bar is 5 $\mu$m.
}\label{fig:FigSx_ABdevice}
\end{figure}

Figure \ref{fig:FigSx_ABdevice} shows a false-colour scanning electron microscope (SEM) image of gate-tunable Device 1 together with an electric circuit diagram indicating the four-terminal measurement configuration. In this device, a NW with width $w\approx$ 150 nm (red) fabricated from a 30-nm-thick BST thin film and capped with Al$_{2}$O$_{3}$ is connected via Pt/Au leads (dark yellow) to the measurement  circuit. A Pt/Au top-gate electrode (green) on top of the transport channel is used to vary the chemical potential. The resistance of the nanowire was measured between the voltage contact pair 4--5, while the current was set to flow from contact 1 to 7.

\section{Gate-voltage dependent resistance oscillations in TINW}

In TINWs, the resistance oscillates as a function of the gate voltage ($V_G$) due to the subband structure (see Fig. \ref{fig:FigSx_ABschematic}b) \cite{Munning2021}; the positions of the minima and maxima relative to the charge-neutrality point (CNP) are set by the perimeter length $L_{\rm P}$ and the capacitance of the device, $C$, as discussed in detail in a previous work \cite{Munning2021}. The spacing of the $V_G$-dependent oscillations is most sensitive to $C$. This $C$ can be reasonably estimated, thanks to the simple slab-like geometry of the thin-film based devices. By using an electrostatic model for our geometry --- including the dielectric environment --- and solving the Laplace equation using the finite element method (in which we make a simple assumption that the NW can be modeled as a perfect metal and that the effects of the quantum capacitance can be neglected \cite{Munning2021}), we obtain a capacitance $C \approx$ 55 pF/m for the parameters of Device 1. This estimated capacitance and the dimensions of the device are used for calculating the expected positions of the resistance peaks (vertical dashed lines) and the theoretically expected oscillations in $R(V_G)$ (solid blue lines) shown in the inset of the main-text Fig. 1g. Note that the peak positions depend on the charge density $n \propto (E_F)^2$, such that the spacing of the $V_G$-dependent oscillations is approximately quadratic in the voltage difference from the location of the CNP. In the main-text Fig. 1g, despite the variations in $L_{\rm P}$ and additional uncertainty in the capacitance, the locations of observed resistance peaks correspond well with the theoretically expected positions, especially for small subband indices $\ell$. 

%


\section{Estimation of the mobility $\mu$ in nanowires}

Since the Hall resistivity cannot be measured on a NW, we tried to estimate the mobility in a NW based on the $R(V_G)$ behavior.  
Note that the total two-dimensional carrier density in a TI thin film due to the surface states is approximately given by $n_{2\mathrm{D}}(k_F)\approx2 \cdot \frac{1}{(2 \pi)^2} \cdot \pi k_F^2=\frac{k_F^2}{2 \pi}$, where $k_F$ is the Fermi wave vector on the Dirac cone.
From the known band structure of BST \cite{Zhang2011}, we know that $k_F \approx 0.06$ \AA$^{-1}$ when the chemical potential is just touching the top of the bulk valence band (BVB). The corresponding hole density in this situation (which we call the threshold hole density $p_{\rm 2D,th}$) is about $6\times10^{-12}$ cm$^{-2}$.

Experimentally, when the chemical potential touches the BVB, the rate of change in $R$ with respect to $V_G$ becomes weaker; this situation corresponds to the region where the slope of $R(V_G)$ starts to flatten after rapidly decreasing with the sweep of $V_G$ to the negative direction. In high-quality BST thin films, a sheet resistance $R_{\square}$ of 2--4 k$\Omega$ is typically observed in this region \cite{Zhang2011, Yang2015, Taskin2017PHE}, which is also the case in our Hall-bar device (see Fig. 1e of the main text); namely, the Hall measurement in our Hall-bar device gives $p_{\mathrm{2D}} \approx 6\times10^{-12}$ cm$^{-2}$ at $V_G$ = $-10$ V where $R_{\square} $= 3.85 k$\Omega$ was observed. From $R_{\square} $= 2--4 k$\Omega$ and  $p_{\rm 2D,th} \approx 6\times10^{-12}$cm$^{-2}$, one obtains the mobility $\mu \simeq$ 250--500 cm$^2$/Vs, which is actually the typical surface-state mobility of a high-quality BST film \cite{Zhang2011, Yang2015, Taskin2017PHE}. 

Based on this knowledge, we tried to make a rough estimate of $\mu$ from the $R(V_G)$ data of the NW shown in Fig. 1g of the main text, where a flattening of the $R(V_G)$ behavior occurs below $V_G \simeq -5$ V.  With $w \approx 150$ nm, $L \approx 650$ nm, and $R \approx$ 5.5 k$\Omega$ at  $V_G = -5$ V, we obtain $R_{\square} \approx$ 1.3 k$\Omega$, which gives $\mu \approx$ 780 cm$^{2}$/Vs if $p_{\rm 2D,th} \approx 6\times10^{-12}$cm$^{-2}$ is assumed as the carrier density. This result suggests that at least no obvious degradation is caused in the process of NW fabrication.


\section{$\pi$-phase shift in the Aharonov-Bohm oscillations}

\begin{figure}[t]
\includegraphics[width=0.95\columnwidth]{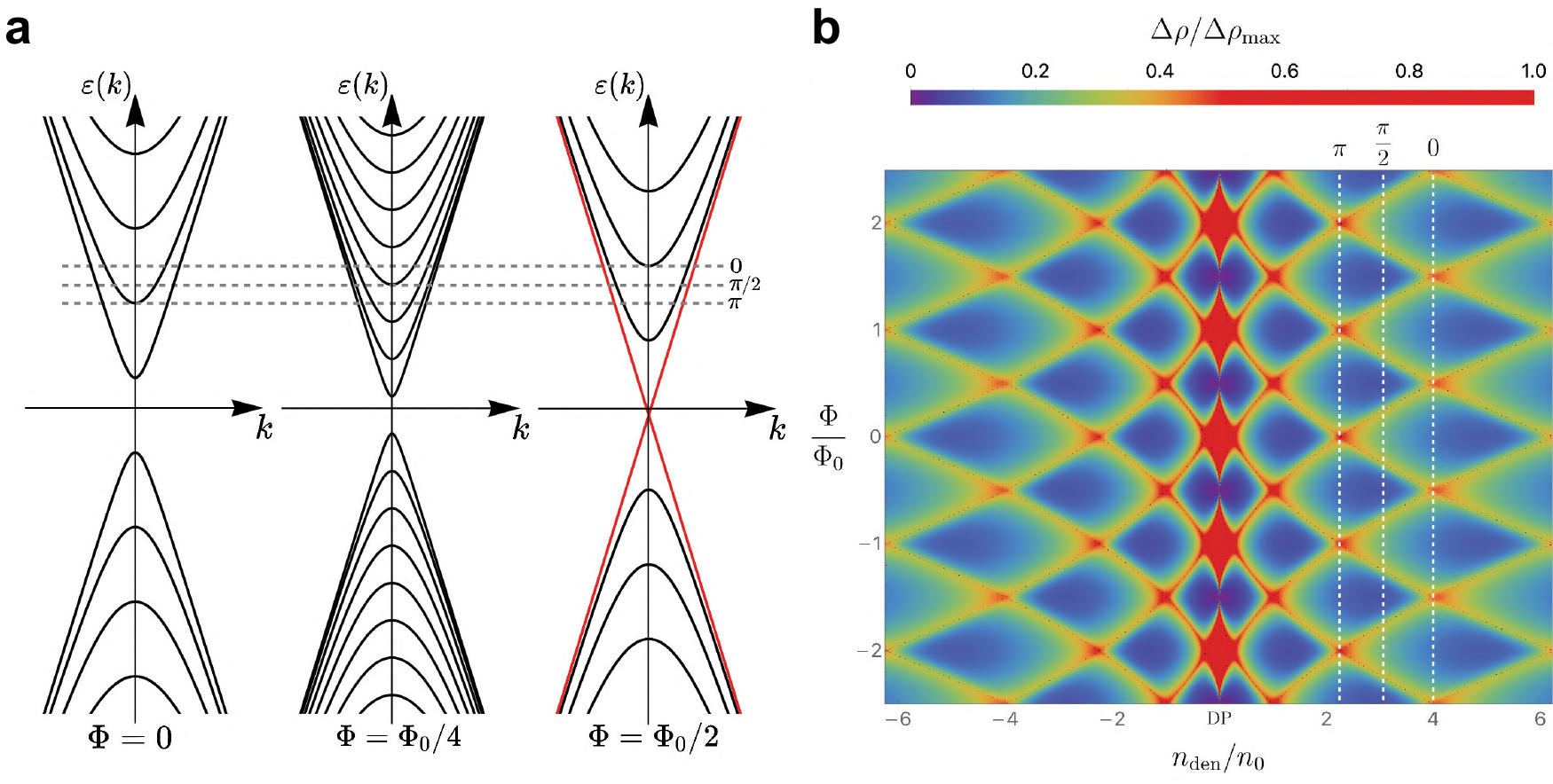}
\caption{\linespread{1.05}\selectfont{} 
\textbf{a,} Schematic of the quantized subbands formed in a TINW for the axial magnetic flux $\Phi$ of 0, $\Phi_0/4$, and $\Phi_0/2$. Red lines indicate the gapless spin-non-degenerate 1D mode. Dashed horizontal lines mark characteristic positions of $E_F$ relevant for a peak in the resistance oscillations.
\textbf{b,} Color mapping of the normalized change in the resistivity, $\Delta \rho / \Delta \rho_{\rm max}$, expected from theoretical calculations as a function of both axial magnetic flux $\Phi$ (normalized by $\Phi_0$) and the carrier density $n_{\rm den}$ near the CNP (normalized by the value $n_0$ at the edge of the first subband). The latter is controlled by gating. White vertical dotted lines denote the three positions of $E_F$ marked in panel {\bf a}. At a fixed $E_F$, $\Delta \rho$ oscillates with $\Phi$ and the oscillation phase shifts depending on $n_{\rm den}$, causing a checker-board-like pattern.
}\label{fig:FigSx_ABschematic}
\end{figure}

The subband structure formed in TINWs due to the quantum confinement causes the density of states to oscillate strongly with energy or with the axial magnetic flux, which manifests itself in terms of resistance oscillations when $E_F$ or $B$ is swept \cite{Zhang2010, Bardarson2010}. In our experiment, both resistance oscillations as a function of $V_G$ (see main-text Fig. 1g) and the Aharonov-Bohm (AB) oscillations as a function of $B$ (see main-text Fig. 2) were observed. In both cases, the oscillations are mainly the result of the increased scattering rate that arises at the edge of each subband (see Fig. \ref{fig:FigSx_ABschematic}a). A characteristic feature of the AB oscillations in TINWs is a $\pi$-phase shift in the oscillations that occurs as $E_F$ is changed \cite{Bardarson2010} (see Fig. \ref{fig:FigSx_ABschematic}b).

\begin{figure}[h]
\includegraphics[width=0.8\columnwidth]{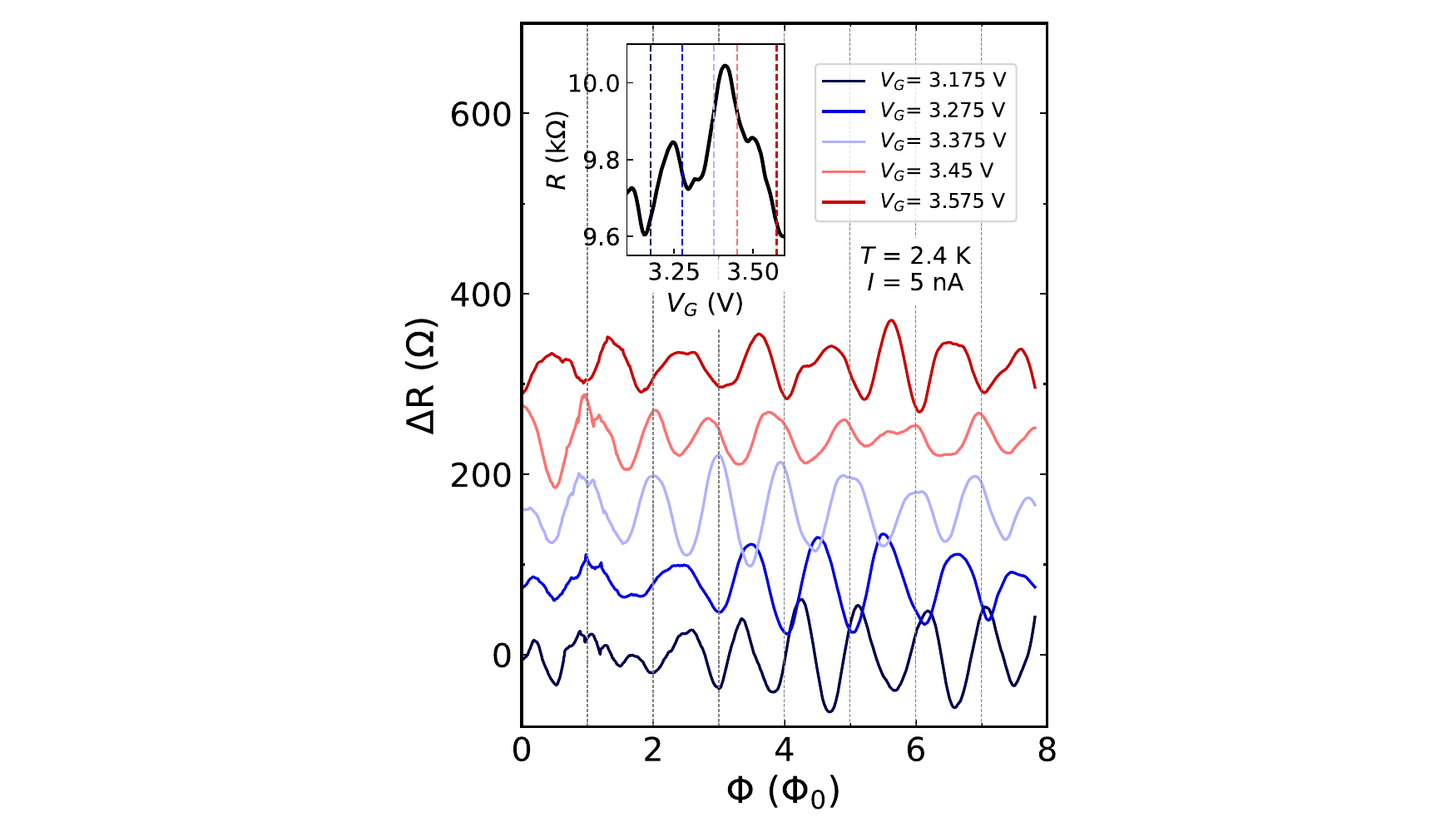}
\caption{\linespread{1.05}\selectfont{}
\textbf{a,} Resistance change $\Delta R$ as a function of magnetic flux $\Phi$ (in units of the flux quantum $\Phi_0$) at selected $V_G$ positions showing $\pi$-phase shifts in the oscillations. Inset shows the positions of the selected $V_G$ values in relation to the $R(V_G)$ behavior near the CNP. 
}\label{fig:FigS5_AB}
\end{figure}

Owing to the bulk-insulating nature and the gate-tunability of our devices, we were able to investigate the occurrence of such a $\pi$-phase shift in the AB oscillations close to the CNP and with respect to features in $R(V_G)$. Figure \ref{fig:FigS5_AB} shows the change in resistance $\Delta R$ plotted vs the threading magnetic flux $\Phi$ (in units of the flux quantum $\Phi_0 = h/e$) close to CNP at selected $V_G$ positions (see inset showing the $R(V_G)$ data in the relevant $V_G$ range). One can recognize $\pi$-phase shifts in the oscillations plotted here, although the phase shift is sometimes smeared due to irregularities. Interestingly, the phase shift is not observed when the CNP (which is located at $V_G$ = 3.41 V) is crossed, which is a theoretically expected behavior \cite{Cook2011,Sacksteder2016}, as one can see in Fig. \ref{fig:FigSx_ABschematic}b.

\section{Properties of the pristine BST films used for the Josephson junction devices}

\begin{figure}[b!]
\includegraphics[width=0.8\columnwidth]{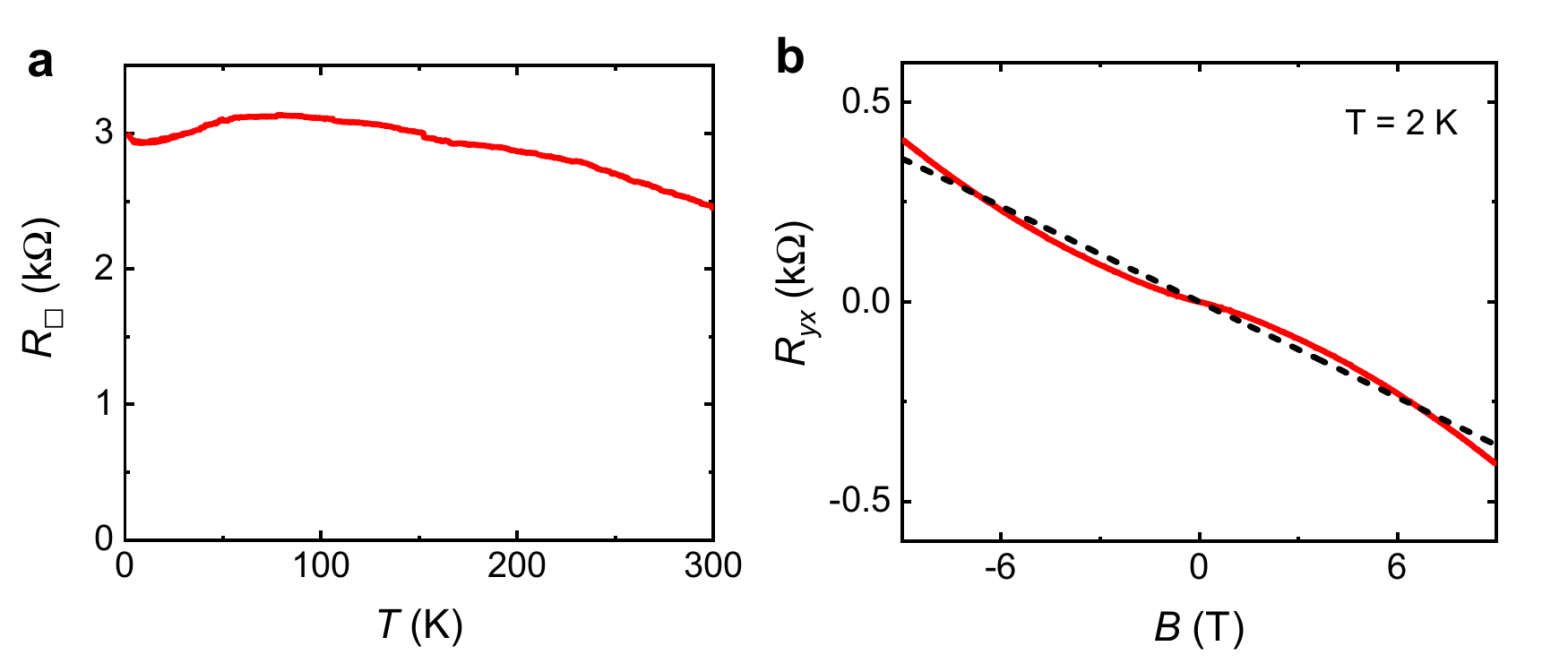}
\caption{\linespread{1.05}\selectfont{}
\textbf{a,} Temperature dependence of the sheet resistance $R_{\square}$ of the pristine BST film used for the Josephson junction 1, whose results are shown in Figs. 3b, 3c, 4a, and 4b of the main text.
\textbf{b,} Hall resistance $R_{yx}$ as a function of magnetic field $B$ of the same thin film.
}\label{fig:FigSx_JJ_BST1}
\end{figure}

Figures \ref{fig:FigSx_JJ_BST1}a and \ref{fig:FigSx_JJ_BST1}b show the $R_{\square}(T)$ behavior and the $R_{yx}(B)$ behavior, respectively, of the pristine BST thin film used for the  Josephson junction 1 before fabrication. A linear fit to $R_{yx}(B)$ gives an apparent carrier density of 1.2$\times10^{13}$ cm$^{-2}$, but this large value is an artifact of the cancellation of the  electron and hole carriers on the two surfaces to give a small $R_{yx}$, which is often observed when the chemical potential is very close the CNP.

\begin{figure}[h]
\includegraphics[width=0.8\columnwidth]{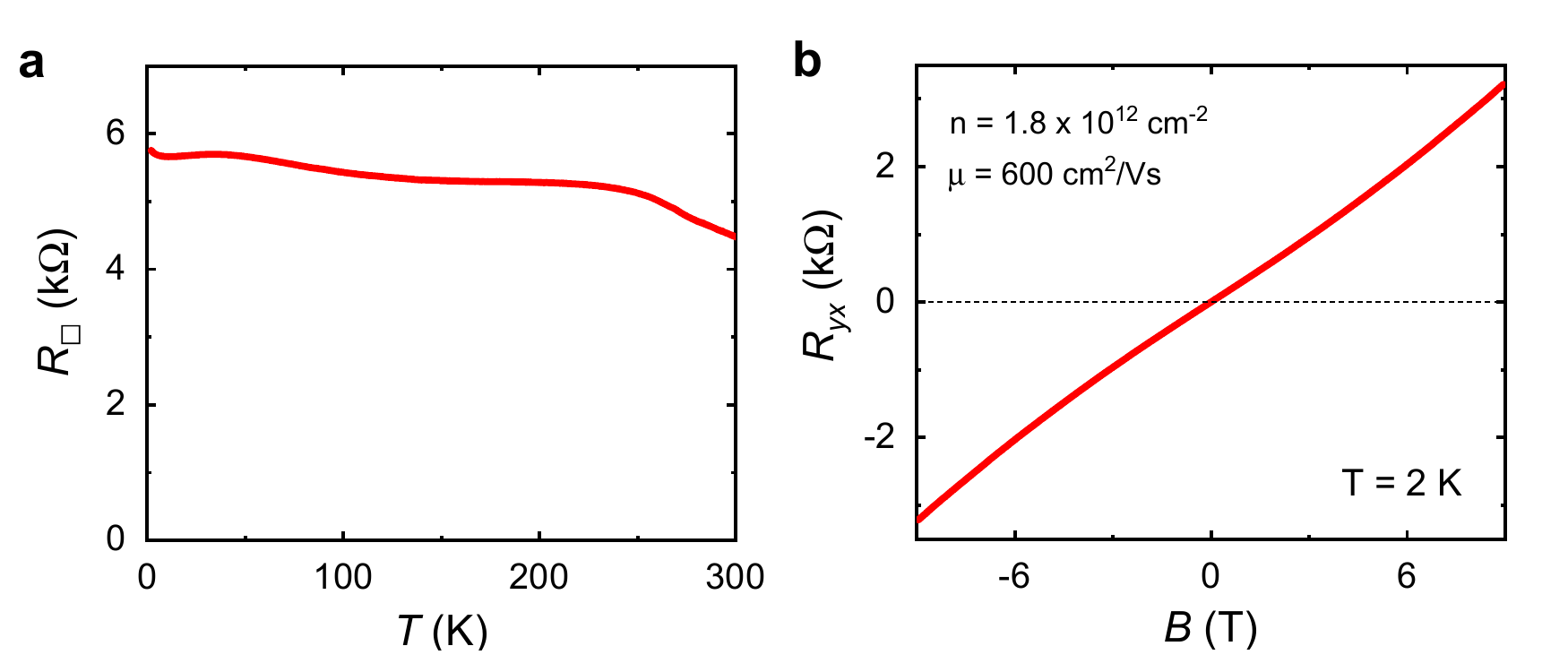}
\caption{\linespread{1.05}\selectfont{}
\textbf{a,} Temperature dependence of the sheet resistance $R_{\square}$ of the pristine BST film used for the Josephson Junctions 2 and 3 used for Figs. 3d, 4c, and 4d of the main text.
\textbf{b,} Hall resistance $R_{yx}$ as a function of magnetic field $B$ of the same thin film.
}\label{fig:FigSx_JJ_BST2}
\end{figure}

Figures \ref{fig:FigSx_JJ_BST2}a and \ref{fig:FigSx_JJ_BST2}b shows the $R_{\square}(T)$ behavior and the $R_{yx}(B)$ behavior, respectively, of the pristine BST thin film used for the  Josephson junction 2 and 3 before fabrication. 
These data indicate that the pristine film was a typical bulk-insulating BST film with $n_{\rm 2D}$ = 1.8$\times10^{12}$ cm$^{-2}$ and $\mu \approx$ 600 cm$^{2}$/Vs.

\section{Temperature dependence of the critical current in Junction 1}

\begin{figure}[h]
\includegraphics[width=0.3\columnwidth]{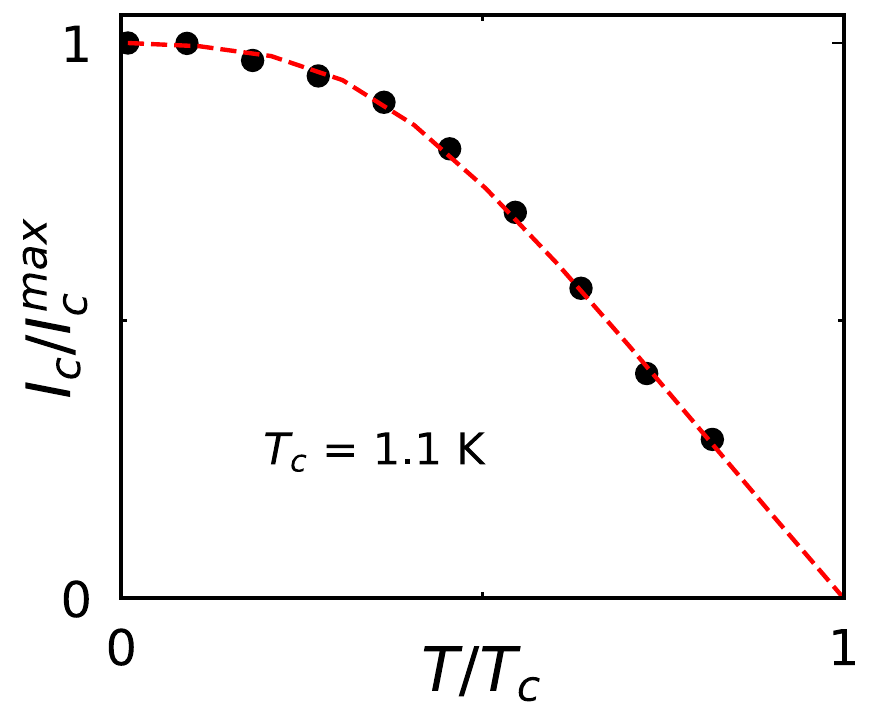}
\caption{\linespread{1.05}\selectfont{}
Temperature dependence of $I_c$ in Junction 1 obtained from d$V$/d$I$ vs $I_{\rm dc}$ at various temperatures up to 0.9 K. The red dashed line is the behavior of the theoretical formula given by Galaktionov and Zaikin \cite{Galaktionov2002} with $T_c \approx$ 1.1 K and transparency $\mathcal{T} \approx$ 0.5.
}\label{fig:JJ1_Ic(T)}
\end{figure}

Figure \ref{fig:JJ1_Ic(T)} shows the temperature dependence of the critical current $I_c$ in Junction 1 ($w$ = 400 nm). Here, $I_c$ was defined as the $I_{\rm dc}$ value at which the junction resistance reaches 10\% of its normal-state resistance $R_N$. Following the analysis of similar TI-based junctions \cite{Bai2022}, we fit the $I_c(T)$ data to the theoretical formula given by Galaktionov and Zaikin \cite{Galaktionov2002}, and the fit yields the interface transparency $\mathcal{T} \approx$ 0.5.

\section{Characteristic parameters of the Josephson-junction devices}

\begin{table}[h]
\setlength{\tabcolsep}{1pt}
\footnotesize
\renewcommand{\arraystretch}{1.1}
\resizebox{\textwidth}{!}{%
\begin{tabular}{|c|c|c|c|c|c|c|c|c|c|c|c|c|c|} \hline
 No. &$L$~(nm)& ~$w$~($\mu$m) & ~$t$~(nm) & ~$I_{\text{c}}$~(nA) & $I_{\text{e}}$~(nA) & ~$R_{\text{N}}$~($\Omega$) & ~$R_{\text{N}}^{I\textrm{-}V}$~($\Omega$) & ~$I_{\text{c}}R_{\text{N}}$~($\mu$V) & ~$ I_{\text{c}}R_{\text{N}}^{I\textrm{-}V}$($\mu$V) & ~$\Delta_{\rm ind}$ (meV) & ~$eI_{\text{c}}R_{\text{N}}/\Delta_{\text{ind}}$ & ~$eI_{\text{c}}R_{\text{N}}^{I\textrm{-}V}/\Delta_{\text{ind}}$ & ~$\mathcal{T}$~ \\  \hline
 1 & 70 & 400 & 14 & 86 & 852 & 238 & 242 & 20.5 & 20.8 & 1.25 & 0.02 & 0.02 & 0.5 \\  \hline
 2 & 65 & 80 & 16 & 4 & 22 & 1770 & 1763 & 7.1 & 7.1 & 1.1 & 0.01 & 0.01 & 0.5 \\  \hline
 3 & 70 & 250 & 16 & 20 & 213 & 518 & 524 & 10.5 & 10.5 & 0.75 & 0.01 & 0.01 & 0.5 \\  \hline
\end{tabular}%
}
\caption{Relevant parameters of the reported Josephson-junction devices.}\label{tab:Table_SC}
\end{table}

\begin{figure}[h]
\includegraphics[width=\columnwidth]{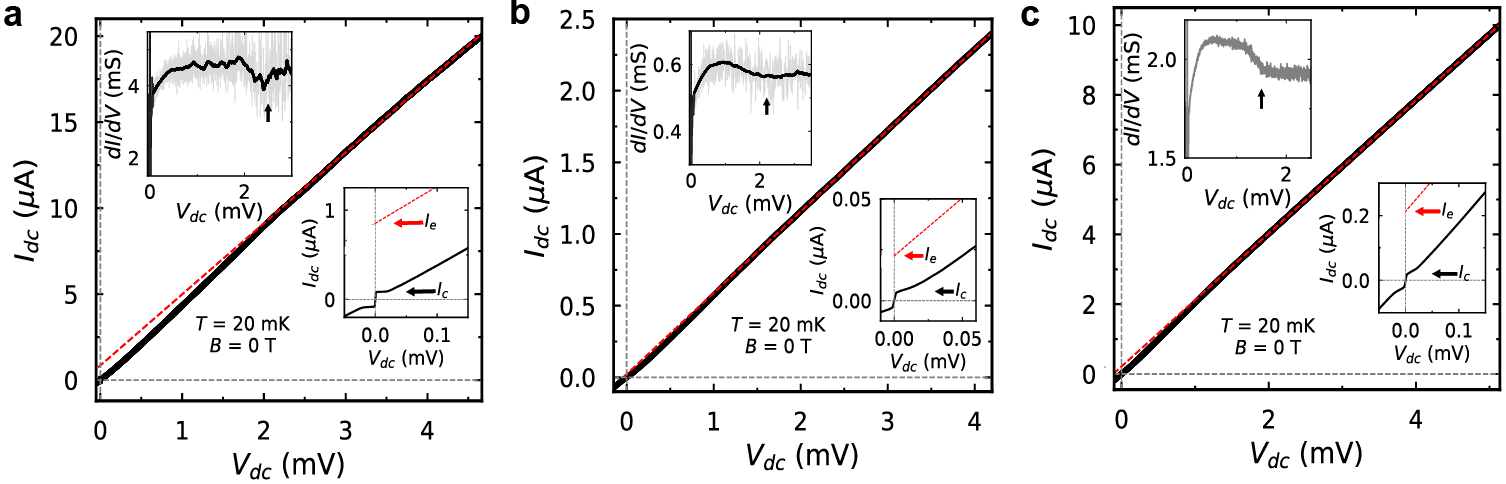}
\caption{\linespread{1.05}\selectfont{}
{\bf a,b,c,} $I$-$V$ characteristics of Junctions 1, 2, and 3, respectively, measured at 20 mK in 0 T. The red dashed line is a linear extrapolation of the high-bias region to $V_{\rm dc}$ = 0 V, used for the extraction of the excess current $I_e$ and the normal-state resistance $R_{N}^{I-V}$. The lower inset in each panel shows a close-up of the main panel at very low $V_{\rm dc}$, with the red arrow indicating $I_e$ and the black arrow indicating $I_c$. The upper inset in each panel shows the differential conductance d$I$/d$V$ as a function of $V_{\rm dc}$; thin grey line shows the raw data and the thick black line is a smoothed curve, with the black arrow indicating the upper limit of $2\Delta_{\rm ind}$/e.
}\label{fig:I-V}
\end{figure}

The characteristic parameters of the three JJs reported in the main text are listed in Table \ref{tab:Table_SC}. $L$ is the channel length of the JJ obtained from the SEM image, $w$ is the width of the TI part, $t$ is its thickness, $I_{\mathrm{c}}$ is the critical current, and $R_{\mathrm{N}}$ is the normal-state resistance measured by applying a large dc current $I_{\mathrm{dc}} \gg I_{\mathrm{c}}$. 

Figure \ref{fig:I-V} shows the $I$-$V$ curves measured on these three JJs.
The excess current $I_{\mathrm{e}}$ is identified from the $I$-$V$ curve by measuring up to a high bias beyond $V_{\mathrm{dc}}=2 \Delta_{\rm Nb} / e$, where $ \Delta_{\mathrm{Nb}} \approx$ 1.3 $\mathrm{meV}$ is the superconducting gap of Nb. 
The linear part of the high-bias $I$-$V$ curve was extrapolated to $V_{\mathrm{dc}}=0$ and the intercept on the $I_{\mathrm{dc}}$ axis gives $I_{\mathrm{e}}$. 
$R_{\mathrm{N}}^{I-V}$ is the inverse slope at high bias (i.e. $V_{\mathrm{dc}}>2 \Delta_{\mathrm{ind}} / e$) region. In low transparency TI-based JJs containing a series of interfaces S-S'-TI-S'-S, the induced gap $\Delta_{\rm ind}$ in S' is the relevant gap for the Andreev reflections at the S'/TI interface that affects the observed $I$-$V$ curve, and the upper limit of $2\Delta_{\rm ind}$ can be inferred from the plot of $dI/dV$ vs $V_{\rm dc}$ as the energy scale above which the $dI/dV$ ceases to change with $V_{\rm dc}$. In the upper inset of each panel in Fig. \ref{fig:I-V}, this energy scale is indicated by an arrow. 
The transparency $\mathcal{T}$ of the interface governing the $I$-$V$ curve can be estimated from the values of $I_{\mathrm{e}}$, $R_{\mathrm{N}}^{I-V}$, and $\Delta_{\rm ind}$ by using the Octavio-Tinkham-Blonder-Klapwijk (OBTK) theory \cite{Octavio1983, Flensberg1988}, and this estimate gives a relatively low transparency of $\mathcal{T} \approx$ 0.5.

\section{Shapiro-step data for Junction 1 at $f$ = 3 GHz}

\begin{figure}[h]
\includegraphics[width=0.7\columnwidth]{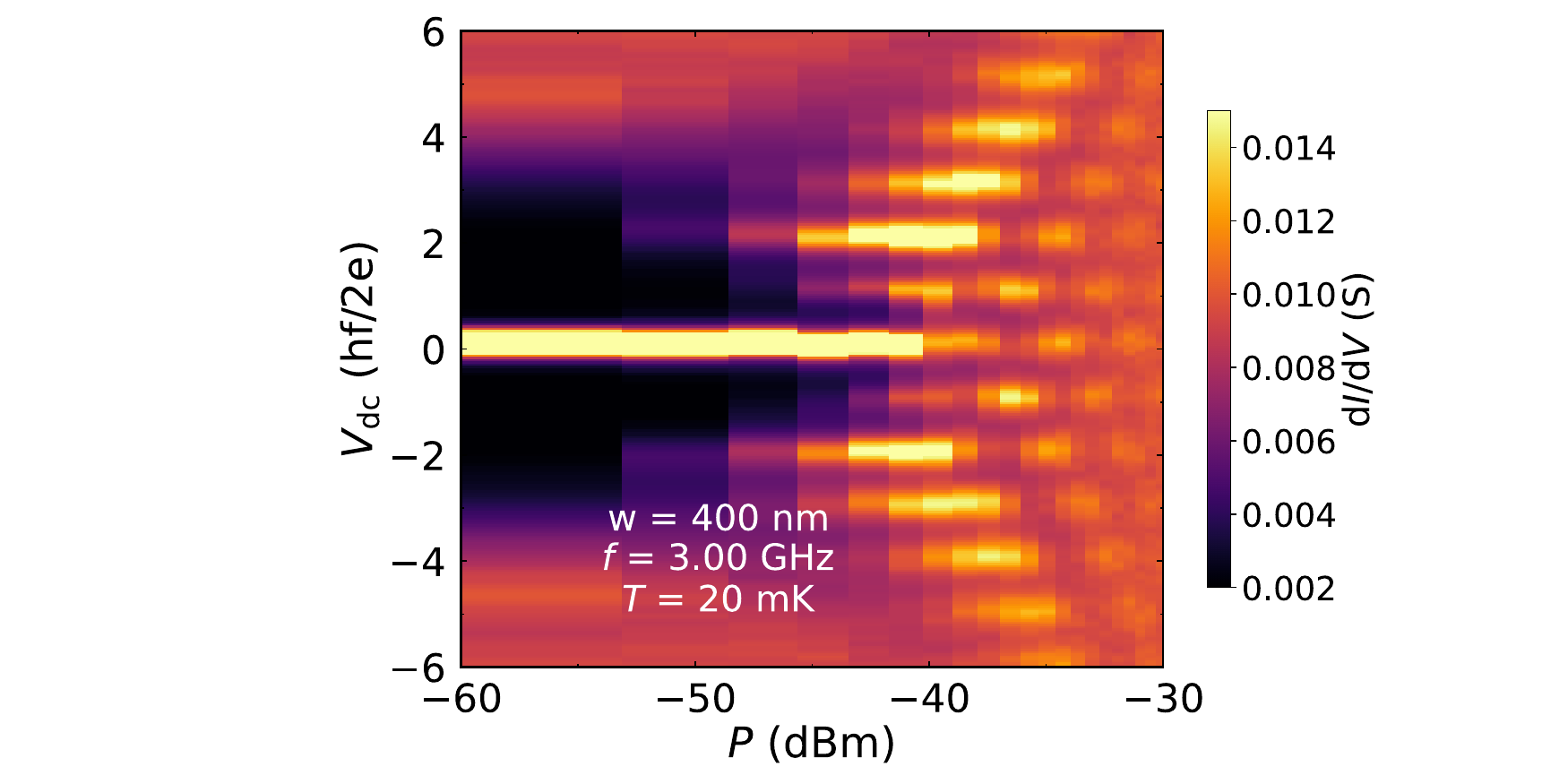}
\caption{\linespread{1.05}\selectfont{}
\textbf{a,} Color mapping of d$I$/d$V$ measured in Junction 1 on a TI ribbon with $w$ = 400 nm under rf irradiation at frequency $f$ = 3 GHz as a function of rf power $P$ and applied dc voltage $V_{\rm dc}$ at $B$ = 0 T and $T$ = 20 mK. 
}\label{fig:FigSx_ShapiroTransition}
\end{figure}

In the main-text Figure 4, we showed the Shapiro-step data only for $f$ = 2.25 and 4 Gz for Junction 1 ($w$ = 400 nm). Figure \ref{fig:FigSx_ShapiroTransition} shows the Shapiro-response of Junction 1 at $f$ = 3 GHz as a function of the rf power $P$ and the bias voltage $V_{\rm dc}$. The 1st Shapiro step is partially observed at relatively high $P$. 

\bibliography{../EtchedTINW.bib}

%
%
%